\documentclass[twoside]{article}
\usepackage{fleqn,espcrc2,epsfig}

\usepackage{graphicx}
\usepackage[figuresright]{rotating}

\newcommand{\AmS}{{\protect\the\textfont2
  A\kern-.1667em\lower.5ex\hbox{M}\kern-.125emS}}

\title{MiniBooNE: Status of the Booster Neutrino Experiment}

\author{Andrew O. Bazarko 
for the MiniBooNE Collaboration 
\thanks{To appear in Proceedings of the XIX International Conference on 
Neutrino Physics and Astrophysics (Neutrino 2000), Sudbury, Canada, 
16-21 June 2000.}
\hfill\break
\hfill\break Department of Physics, 
Princeton University, Princeton, New Jersey 08544-0708, USA
}
\begin{document}

\begin{abstract}
MiniBooNE is preparing to search for $\nu_\mu\rightarrow\nu_e$ oscillations
at Fermilab.  The experiment is designed to make a conclusive statement
about LSND's neutrino oscillation evidence.  We give a status report 
on the preparation of the experiment and outline the 
experimental prospects. 
\vspace{1pc}
\end{abstract}

\maketitle

\section{INTRODUCTION}

There are now three reported indications for neutrino oscillations. 
The most compelling evidence comes from experiments studying 
neutrinos produced by cosmic rays in the earth's atmosphere, which
find that muon neutrinos $(\nu_\mu)$ 
entering the detector from below (which had to travel through the earth) are depleted  
when compared with those incident from above (which traveled a shorter distance) \cite{atmos}.
There is also evidence from solar neutrino experiments, which observe fewer electron
neutrinos $(\nu_e)$ than would be consistent with the sun's energy output \cite{solar}.
Finally, there is evidence from a lone accelerator experiment, LSND at Los Alamos, 
which observed an excess of $\bar\nu_e$ events 
from a predominantly $\bar\nu_\mu$ beam \cite{lsnd}. 

The full picture of neutrino oscillations remains 
incomplete.  The three reported indications differ
according to amplitude, frequency, and what neutrino flavors participate, and  
cannot be explained by oscillations involving 
only the three known neutrino flavors.  Perhaps further study will 
reveal a fourth neutrino.

MiniBooNE is in a unique position to help address this possibility, 
because MiniBooNE is alone among upcoming experiments with sensitivity 
to the LSND mode of oscillations.  MiniBooNE is designed to 
confirm or refute LSND unequivocally, and thereby either keep alive or kill evidence
for a fourth neutrino.  The MiniBooNE Collaboration is 
now building a new beam and detector at Fermilab, and we plan to start collecting 
data in December 2001. 

%\subsection{Spacing}

\section{EXPERIMENTAL OVERVIEW}

At MiniBooNE we will search for the oscillation of muon neutrinos into 
electron neutrinos $(\nu_\mu \rightarrow \nu_e)$.  
We will produce a $\nu_\mu$ beam in the energy range
0.5 -- 1.0 GeV with a small intrinsic
$\nu_e$ component (less than 0.3\%) and search for an excess of
electron neutrino events in a detector located approximately 500 m from the neutrino source.  
The baseline to neutrino energy ratio will thereby be similar to that of 
LSND, $L/E \sim 1$, giving MiniBooNE sensitivity to the same mode of
oscillations.  However, neutrino energies will be more than an order 
of magnitude higher than at LSND, so that the search at MiniBooNE will 
be in a different environment and will 
employ different experimental signatures.

The MiniBooNE neutrino beam will be initiated by a primary beam of 8 GeV protons
from the Fermilab Booster.  The Booster is a reliable, high intensity 
machine, expected to run at least $2\times 10^7$ s per year, 
while delivering $5\times 10^{12}$ protons per 1.6 $\mu$s pulse
at a rate of 5 Hz to MiniBooNE.  One year for 
the experiment is expected to correspond to $5\times10^{20}$ protons on target.
The Booster has the capacity to provide protons for  
several Fermilab efforts,  
%MiniBooNE will run alongside the other Fermilab programs.
and when MiniBooNE starts, the Booster will supply beam to 
both the Tevatron collider 
and to MiniBooNE.  Later, we anticipate that the Tevatron, NuMI, and
MiniBooNE programs can be accommodated simultaneously. 

\begin{figure}[tb]
\vspace{9pt}
\includegraphics[width=75mm]{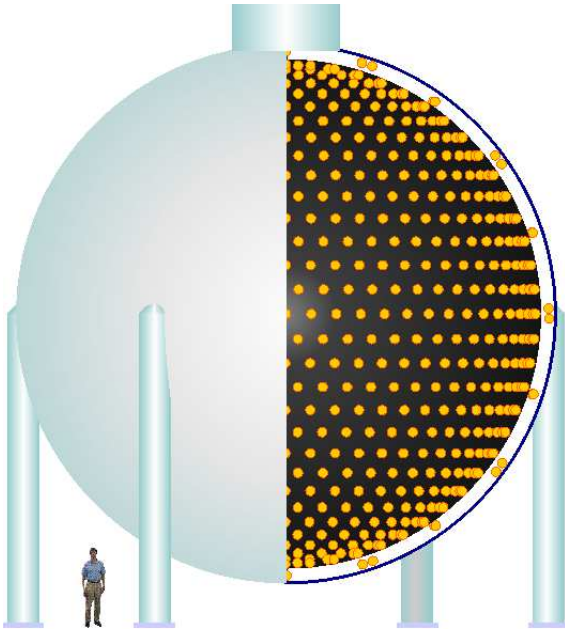}
\caption{A schematic of the MiniBooNE detector.  The cutaway shows the layout of 
8-inch phototubes in the black main region, and in the white veto region.} 
\label{fig:detector}
\vspace*{-8mm}
\end{figure}

A secondary beam will be produced when 
the 8 GeV protons strike a beryllium target positioned inside  
a magnetic horn.  
Positively charged particles (mostly pions) from  
the target will be focused forward by the single horn into a 50 m decay channel.
Decays of these secondaries will constitute the neutrino beam. 
Decay lengths of 50 and 25 m will be possible through the use of 
two steel and concrete beam absorbers.  
One absorber will be permanently positioned at the end of the decay channel.
The intermediate absorber is designed either to be lowered into the 
decay channel or to be raised out of the way.  
This ability to vary the decay length will provide a check of experimental 
systematics associated with $\nu_e$ contamination.
% from the decay of muons in the secondary beam. 

The MiniBooNE neutrino 
detector will consist of 800 tons of pure mineral oil contained in a 40-foot (12.2 m) 
diameter spherical tank.  A structure in the tank will support phototubes, which will
detect neutrino interactions in the oil by the Cherenkov and scintillation light 
that they produce.  (The undoped mineral oil will 
scintillate modestly from the presence of intrinsic impurities.) 
The phototube support structure will also optically isolate the most outer 35 cm 
of oil from the rest, turning the outer oil into a veto region that should stay quiet
while a neutrino produces light only in the inner, main region.
The main region will be viewed by 1280 8-inch phototubes, providing
10\% photocathode coverage of the 445 ton fiducial volume.
The veto region will contain 240 8-inch phototubes mounted in pairs on the tank wall.  
In order to limit reflections, main volume surfaces 
will be black in color.  Surfaces in the veto region,
including the tank wall, will be white, in order to maximize light collection.
A schematic of the detector is shown in Fig. \ref{fig:detector}.

\begin{figure}[tbh]
\vspace{9pt}
\includegraphics[width=75mm]{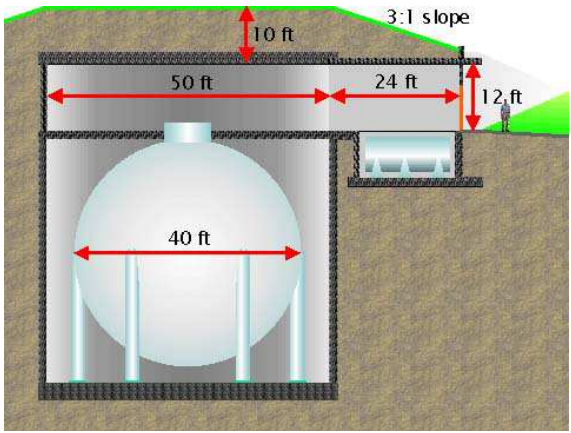}
\caption{Cutaway of the underground MiniBooNE detector enclosure. 
An oil overflow tank is shown underneath the corridor.}
\label{fig:cutaway}
\end{figure}

The center of the detector will be positioned about 490 m from 
the end of the decay channel, and about 6 m below
ground, corresponding to the level of the neutrino beam.  
The tank will be housed in a cylindrical concrete vault.
A room above the vault will contain the experiment's electronics.  
Approximately 3 m of soil will be placed on top of the 
enclosure with access through a corridor, as shown in Fig.
\ref{fig:cutaway}.

\section{STATUS}

A few of MiniBooNE's features have changed from those described in
the proposal \cite{proposal}. Most notable is the switch from 
two-horn secondary beam focusing to a single horn design. 

Much work is ongoing to allow MiniBooNE to start running
in December 2001.  The status on several fronts as of August 2000 
is reported here.

\subsection{Civil construction} 

Construction of the detector enclosure is nearing completion.  
Pieces of the steel tank arrived at Fermilab in early April and were
assembled inside the detector vault.  In June, the completed
tank was successfully hydrotested, that is, it was filled with water, 
subjected to a slight overpressure, and found not to leak. 
Since then, the electronics room has been completed and the earth
berm up to the roof of the enclosure has been constructed.   
The tank is being painted and the enclosure is being outfitted with utilities.
We experimenters expect to take occupancy of the space in October. 

Construction of the 8 GeV beam, the target hall, and the decay channel began in July.  
Concrete at the foundation of the target hall is in place. The target hall walls 
are to go up next, and an opening in one wall will lead to the decay channel. 
The decay channel will be enclosed in a waterproof liner, to prevent 
migration of radioactivity from the soil near the channel.  The liner 
material remains pliable down to a few degrees Celsius, so it cannot be installed 
during the cold of winter.  Therefore, an important item on the beam construction
project schedule is liner installation by late October, which appears feasible.  

\subsection{Detector components} 

MiniBooNE will reuse the phototubes from LSND together with 
330 new tubes acquired from Hamamatsu. 
LSND's 1220 phototubes are now at Fermilab, and shipments of
the new tubes have been arriving over the summer.  Both new and
old tubes are being tested to characterize charge and time
resolutions and to determine operating voltages.  
Results for the LSND tubes correspond nicely with old
measurements, indicating that these old tubes remain stable after
their journey from New Mexico to Illinois.  Tests also verify the better
performance of the new tubes, which have a greater number of dynode stages and
narrower time resolution. 

A prototype of the phototube support structure has been assembled and
final components are being fabricated.  
The bulk of this aluminum structure is being painted, 
black on one side (to face the main volume) and white on the other (to face
the veto).  
Aircraft paints have been chosen.  
The white offers high albedo over most of the visible 
spectrum, whereas the black is quite non-reflective, even at large angles of 
incidence. These paints offer good adhesion to aluminum and our tests indicate that 
they will not contaminate the mineral oil.  Some smaller aluminum pieces used to mount 
phototubes in the main region will be sandblasted and black anodized, 
a process that also meets our low-albedo and low-contamination requirements. 

Preparation of the cables that will connect the phototubes to pre-amplifier cards outside
the tank has started. 
Each phototube will be attached to one cable that will both supply the high voltage 
and carry out the signal. Cables will be routed along the wall of the tank, and they
will penetrate individually into the main volume as needed. 
Because these cables will run through the veto region, they will be jacketed in white, 
and teflon jacket material has been chosen because it 
is non-contaminating (our tests indicate that 
standard PVC cable jackets contaminate mineral oil).
Each phototube will have a short ``pigtail'' of black cable to which the white cable 
will be spliced.  Thus black cable will be in the main 
volume and white cable will be in the veto region. 
 
Work is ongoing on other parts of the data acquisition system.
Along with the phototubes, the charge and time readout electronics from LSND, 
the so-called QT cards and crates, will be reused.  
New high voltage power supplies are being acquired, 
and new event building hardware and software are being developed.
Progress is also being made on calibration systems.  A laser
calibration system will provide short pulses of light from a tunable dye laser
to four light-diffusing flasks at various locations in the detector.  This system is similar
to one used successfully in LSND to determine phototube time offsets and time slewing
corrections.  In addition, a cosmic ray calibration system is being built.  This system 
will consist of a muon tracker (scintillator hodoscope) 
above the detector with scintillator cubes (5 cm on a side)  
located under the tracker inside the detector.  The system will provide 
the entering position and direction of cosmic ray muons that stop in the 
detector, including the fraction that stop in a cube.  These events will be 
used to calibrate the position, energy, and direction determination of the 
reconstruction algorithm. 

A scaffold will be constructed inside the tank for installation of the 
phototubes and the other detector internals.
Phototube installation is 
expected to take four to five months and should begin in January 2001.
Filling the tank with mineral oil is planned for fall 2001. 

\subsection{Beam components} 

An array of components --- magnets, power supplies,
vacuum systems, water cooling systems, beam monitors, collimators, etc. ---
is being prepared to transport the 8 GeV protons from the Booster to the target hall 
and form the neutrino beam. 
The most challenging element to construct is the magnetic horn 
that will focus the secondary beam.  

The MiniBooNE horn is designed to operate with 170 kA pulses at a rate of 5 Hz, 
with the goal that it withstand 200 Mpulses (two years of operation).
This is a higher repetition rate and more pulses than any previous horn.
The horn is designed to tolerate the stresses that each pulse will bring
as the current heats the conductors and generates magnetic forces.  
The heat will be managed with water cooling, and the horn body is designed to 
contain the radioactive water that will circulate inside. 
Numerous stress calculations have been performed, with concern  
over such issues as metal fatigue, structural resonances, and water erosion.
In the end, however, tests on a prototype will settle the design.

Several horn and target pre-prototype tests have been completed. 
For example, cooling tests for the 65 cm air-cooled beryllium target 
were completed, and target construction is
on track.  Various elements of the horn and its power supply have been tested, 
such as the water cooling system, the welding procedures, and the possibility 
of excessive corona currents in the design of high potential difference areas.
Parts for a prototype horn are being procured, and a proto-horn will be 
assembled over the winter, with tests scheduled for spring 2001.  The plan is
to test the proto-horn for at least 10 Mpulses, which will mark an important  
milestone in the beamline construction.  Ultimately, we plan to have an installed
horn and a spare, in addition to the proto-horn. 

The target and horn will be contained under a pile of steel and concrete shielding in 
the target hall. ``Hot horn handling'' procedures 
are being planned to limit exposures to radioactivity should horn replacement be necessary.  
These involve putting the radioactive target and horn assembly into a ``coffin''
for safe removal from the target hall. 

\section{OUTLINE OF FUTURE ANALYSIS}

We will reconstruct quasielastic $\nu_e$ 
interactions by identifying electrons via their characteristic
Cherenkov and scintillation light signatures. 
Besides the $\nu_e \rightarrow e^-$ signal, several backgrounds 
will contribute.  The analysis will come down to 
accounting for the backgrounds and determining whether or not 
there is an excess.
The background sources will be due to  $\nu_e$ contamination in the beam
and to the misidentification as electrons of 
muons and $\pi^0$'s produced in the detector.  
Because the neutrinos are at higher energies than at LSND, 
neutrons will not play a role in the signal and will not contribute
background.  

The detector will record the time of the initial hit and total charge 
for each phototube.   From this information, the track position
and direction will be determined.  Muon tracks will be distinguished 
from electron tracks by their Cherenkov rings and  
scintillation light.  Electrons will
tend to produce ``fuzzy'' rings due to multiple scattering and 
bremsstrahlung, while muon rings will tend to have sharp outer boundaries.
Electrons also tend to have a high fraction of prompt (Cherenkov)
light compared to late (scintillation) light, whereas muons 
produce relatively more late light.

MiniBooNE's secondary beam will be dominated by pions $(\pi^+)$ with energies around 2 GeV. 
Pions decay to a $\nu_\mu$ and a muon 99.988\% of the time.  Most of the muons $(\mu^+)$, because of 
their relatively long lifetime, will travel to the end of the decay channel and stop in the 
absorber or the dirt.  
Any muon that decays, however, will yield a $\nu_e$.
So, pions are an almost pure source of $\nu_\mu$'s, but the muons
that are produced along with them contribute some $\nu_e$ contamination.  
This situation is not as bad as it might seem, because 
both the muon and the $\nu_\mu$ emerge from the same two-body pion decay, which allows the 
measurement of one of the decay particles to constrain the other. 
In one year of running, the MiniBooNE detector will measure about 0.5 million $\nu_\mu$ 
events.  These $\nu_\mu$ events will act as excellent monitors of the pions that produced
them, because the two-body kinematics, relatively low beam energy, and the small 
angle subtended by the detector conspire to produce a very tight
correlation between $\pi^+$ and $\nu_\mu$ energies.  Thus, the pion secondary beam will be  
well constrained, and then so too will be the $\pi^+\rightarrow\mu^+\rightarrow\nu_e$ chain.
In this way, we expect to reduce the uncertainty in the muon component of the $\nu_e$ background
to less than 5\%.  

While pions will comprise most of the secondary beam, some kaons will also be produced
at the target.  Charged kaons decay via $K^+\rightarrow\pi^0 e^+\nu_e$  4.8\% of the time, 
and neutral kaons decay via $K_L\rightarrow\pi^\pm e^\mp\nu_e$ 38.8\% of the time. 
With many fewer kaons than pions produced, the overall kaon contribution to $\nu_e$ 
contamination is expected to be about one third of that due to the muons.  
Because the $K^+$'s
are focused by the horn and the $K_L$'s are not, the $K^+$ $\nu_e$ component will be about twice
that of the $K_L$.  We plan to use Monte Carlo simulation
constrained by production data to limit the systematic
uncertainty in the kaon component of the $\nu_e$ background to about 10\%.
We continue to explore ways to directly measure the kaon content of our secondary beam.

The number of muons that will be misidentified as electrons by the reconstruction 
algorithm can be estimated using the data. 
Ninety two percent of the muons contained in the detector will 
decay, and they will be relatively easily identified by the presence of 
a second track (the decay electron).  However, the 8\% of muons that get captured
have a greater chance to be misidentified.  
The misidentification of muon captures 
will be estimated by studying 
the large sample of muons that decay and determining the  
particle identification algorithm performance while ignoring the 
decay track.  Using this technique, which does not
rely on Monte Carlo simulation, the 
muon misidentification uncertainty is expected to be below 5\%.

Most neutral pions will be identified by their two electromagnetic 
decay tracks.  A small fraction (1\%) of asymmetric $\pi^0$ decays
will not yield two resolvable tracks and will therefore be more likely to
be misidentified. 
The misidentification contribution of these decays will be studied 
with Monte Carlo simulation, which will be constrained by 
the large sample of measured $\pi^0$'s in the experiment. 
The pion misidentification uncertainty is expected to be 5\%.  

\section{PROSPECTS}

If oscillations occur as indicated by LSND, 
MiniBooNE will observe an excess of several hundred electron events 
in one year of running. 
The significance of this excess would be on the 
order of 8 to 10 $\sigma$ above background expectations.

\begin{figure}[t]
\vspace{9pt}
\includegraphics[bb=24 155 525 645,width=75mm]{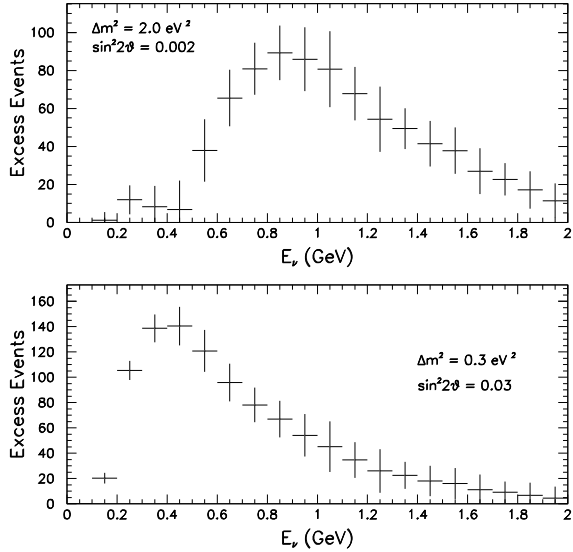}
\caption{Energy spectra of the $\nu_e$ event excess expected for two oscillation scenarios. 
The excess for $\Delta m^2=2$ eV$^2$, $\sin^2 2\theta=0.002$ is shown in the upper plot
and that for $\Delta m^2=0.3$ eV$^2$, $\sin^2 2\theta=0.03$ is in the lower plot. 
The error bars
include both statistical and systematic uncertainties, the latter of which are correlated 
from bin to bin.}
\label{excess}
\end{figure}

In addition to a count of excess events, we will be able
to measure their energy distribution, and thereby enhance 
the experimental sensitivity. 
The expected energy spectra for excess events in
two oscillation scenarios are shown in Fig. \ref{excess}.
Backgrounds are expected to have 
different energy distributions from the oscillation events, so  
an underestimate of the background will not
necessarily be interpreted as a fictitious oscillation signal.

\begin{figure}[h]
\vspace{9pt}
%\framebox[55mm]{\rule[-21mm]{0mm}{43mm}}
\includegraphics[bb=100 150 510 680,width=75mm]{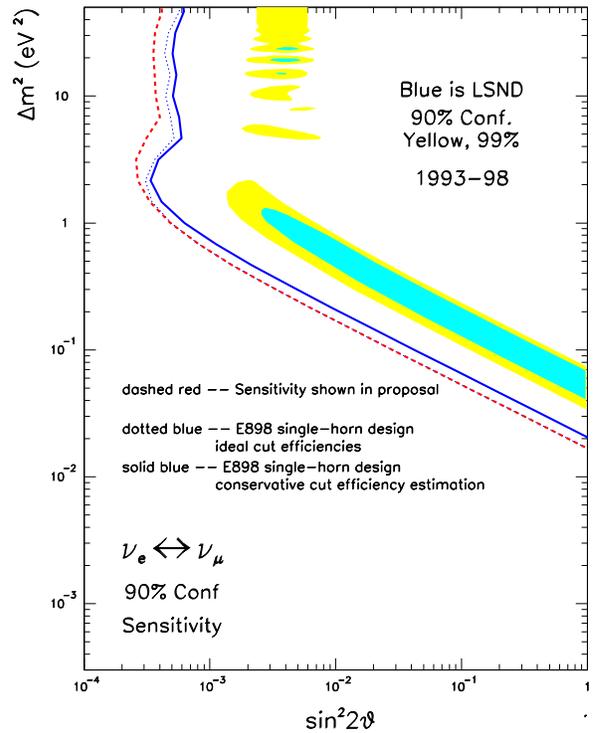}
\caption{
MiniBooNE expected 90\% confidence level sensitivity limits.
The solid curve is the expectation for the final single horn design 
and an analysis with conservative efficiency assumptions.  The dotted
curve is for the single horn and ideal efficiency.  The dashed curve
is the sensitivity using a two-horn design as described in the 
experiment's proposal.}
\label{fig:boone}
\end{figure}

Fig. \ref{fig:boone} shows MiniBooNE's expected 
exclusion contours.

\end{document}